\documentstyle[]{mn}
\newif\ifAMStwofonts

\def\etal{{\rm et al.}}

\def\simgt{\mathrel{\spose{\lower 3pt\hbox{$\sim$}}
        \raise 2.0pt\hbox{$>$}}}
\def\simlt{\mathrel{\spose{\lower 3pt\hbox{$\sim$}}\raise 2.0pt\hbox{$<$}}}

\ifoldfss
  \newcommand{\rmn}[1] {{\rm #1}}

  \ifCUPmtlplainloaded \else
    \NewTextAlphabet{textbfit} {cmbxti10} {}
    \NewTextAlphabet{textbfss} {cmssbx10} {}
    \NewMathAlphabet{mathbfit} {cmbxti10} {} 
    \NewMathAlphabet{mathbfss} {cmssbx10} {} 
  \fi
  \ifAMStwofonts
    \ifCUPmtlplainloaded \else
      \NewSymbolFont{upmath} {eurm10}
      \NewSymbolFont{AMSa} {msam10}
      \NewMathSymbol{\upi}     {0}{upmath}{19}
      \NewMathSymbol{\umu}     {0}{upmath}{16}
      \NewMathSymbol{\upartial}{0}{upmath}{40}
      \NewMathSymbol{\leqslant}{3}{AMSa}{36}
      \NewMathSymbol{\geqslant}{3}{AMSa}{3E}

    \fi
  \fi
\fi 

\ifnfssone
  \newmathalphabet{\mathit}
  \addtoversion{normal}{\mathit}{cmr}{m}{it}
  \addtoversion{bold}{\mathit}{cmr}{bx}{it}
  \newcommand{\rmn}[1] {\mathrm{#1}}

  \newmathalphabet{\mathbfit} 
  \addtoversion{normal}{\mathbfit}{cmr}{bx}{it}
  \addtoversion{bold}{\mathbfit}{cmr}{bx}{it}
  \newmathalphabet{\mathbfss} 
  \addtoversion{normal}{\mathbfss}{cmss}{bx}{n}
  \addtoversion{bold}{\mathbfss}{cmss}{bx}{n}
  \ifAMStwofonts
    \ifCUPmtlplainloaded \else
      \UseAMStwoboldmath
      \makeatletter
      \new@mathgroup\upmath@group
      \define@mathgroup\mv@normal\upmath@group{eur}{m}{n}
      \define@mathgroup\mv@bold\upmath@group{eur}{b}{n}
      \edef\UPM{\hexnumber\upmath@group}
      \new@mathgroup\amsa@group
      \define@mathgroup\mv@normal\amsa@group{msa}{m}{n}
      \define@mathgroup\mv@bold\amsa@group{msa}{m}{n}
      \edef\AMSa{\hexnumber\amsa@group}
      \makeatother
      \mathchardef\upi="0\UPM19
      \mathchardef\umu="0\UPM16
      \mathchardef\upartial="0\UPM40
      \mathchardef\leqslant="3\AMSa36
      \mathchardef\geqslant="3\AMSa3E
    \fi
  \fi
\fi 

\ifnfsstwo
  \newcommand{\rmn}[1] {\mathrm{#1}}

  \DeclareMathAlphabet{\mathbfit}{OT1}{cmr}{bx}{it}
  \SetMathAlphabet\mathbfit{bold}{OT1}{cmr}{bx}{it}
  \DeclareMathAlphabet{\mathbfss}{OT1}{cmss}{bx}{n}
  \SetMathAlphabet\mathbfss{bold}{OT1}{cmss}{bx}{n}
  \ifAMStwofonts
    \ifCUPmtlplainloaded \else
      \DeclareSymbolFont{UPM}{U}{eur}{m}{n}
      \SetSymbolFont{UPM}{bold}{U}{eur}{b}{n}
      \DeclareSymbolFont{AMSa}{U}{msa}{m}{n}
      \DeclareMathSymbol{\upi}{0}{UPM}{"19}
      \DeclareMathSymbol{\umu}{0}{UPM}{"16}
      \DeclareMathSymbol{\upartial}{0}{UPM}{"40}
      \DeclareMathSymbol{\leqslant}{3}{AMSa}{"36}
      \DeclareMathSymbol{\geqslant}{3}{AMSa}{"3E}
    \fi
  \fi
\fi 

\ifCUPmtlplainloaded \else
  \ifAMStwofonts \else 
    \def\upi{\pi}
    \def\umu{\mu}
    \def\upartial{\partial}
  \fi
\fi

\title[The size of the mid-IR emission region in Q2237+0305]
  {The size of a quasar's mid-IR emission region inferred from microlensed images of Q2237+0305}
\author[J. S. B. Wyithe et al.]
  {J.~S.~B.~Wyithe$^{1,2}$, 
  E.~Agol$^3$,
  C. J.~Fluke$^4$\\
  $^1$ School of Physics, University of Melbourne, Parkville, Vic, 3052, 
Australia\\
  $^2$ Princeton University Observatory, Peyton Hall, Princeton, NJ 08544, USA\\
  $^3$ Physics and Astronomy Department, Johns Hopkins University, Baltimore, MD 21218, USA\\
  $^4$ Centre for Astrophysics and Supercomputing, Swinburne University of Technology, Hawthorn, Victoria 3122 Australia\\
 Email: swyithe@physics.unimelb.edu.au, agol@pha.jhu.edu, cfluke@swin.edu.au}
\date{Accepted Received}
\pagerange{\pageref{firstpage}--\pageref{lastpage}}
\pubyear{1999}

\def\LaTeX{L\kern-.36em\raise.3ex\hbox{a}\kern-.15em
    T\kern-.1667em\lower.7ex\hbox{E}\kern-.125emX}

\begin{document}

\label{firstpage}

\maketitle

\begin{abstract}
We use published mid-IR and V-band flux ratios for images A and B of Q2237+0305 obtained in September 1999 to demonstrate that the size of the mid-IR emission region has a scale comparable to or larger than the microlens Einstein Radius (ER) ($\sim 10^{17}$cm for microlensing by solar mass stars). Q2237+0305 has been monitored extensively in the R and V-bands for $\sim$15 years. The variability record shows significant microlensing variability of the optical emission region, and has been used by several studies to demonstrate that the optical emission region is much smaller than the ER for solar-mass objects. For the majority of the monitoring history, the optical flux ratios have differed significantly from those predicted by macro-models. In contrast, recent observations in mid-IR show flux ratios similar to those measured in the radio, and to predictions of some lens models, implying that the mid-IR flux is emitted from a region that is at least 2 orders of magnitude larger than the optical emission region. We have calculated the likeli-hood of the observed mid-IR flux ratio as a function of mid-IR source size given the observed V-band flux ratio. The expected flux ratio for a source having dimensions of $\sim$1 ER is a sensitive function of the macro model adopted. However we find that the probability of source size given the observed flux ratios is primarily sensitive to the ratio of the macro-model magnifications. Limits on the mid-IR source size can therefore be considered as a function of a one dimensional, rather than a 4 dimensional (two optical depths plus two shears) class of models. The majority of published macro models for Q2237+0305 yield a flux ratio for images B and A of 0.8 - 1.1. By combining probabilities from the ratios A/B and C/D we infer that the diameter of a circular IR emission region is $>1$ER with $>95\%$ confidence. For sources of this size, other geometries, specifically an annular geometry, appropriate for a dusty torus, yield the same limit if the projected area rather than radius is considered. For microlensing by low-mass stars, this source size limit rules out non-thermal processes such as synchrotron as mechanisms for mid-IR emission.

\end{abstract}


\section{Introduction}

Q2237+0305 (The Einstein Cross) was discovered in the CfA Redshift survey (Huchra et al. 1985). The object comprises a source quasar with a redshift of $z=1.695$ that is gravitationally lensed by a foreground galaxy ($z=0.0394$) producing 4 images with separations of $\sim 1''$. As a result of the proximity of the lensing galaxy, Q2237+0305 provides a unique opportunity to contrast dynamical measurements based on the geometry of the lensing with more traditional techniques (e.g. Rix, Schneider \& Bahcall 1992, hereafter RSB92). In addition, the close proximity results in a short time-scale for microlensing, and a large projected microlensing length-scale (with respect to the source). Each of the 4 images are observed through the bulge of a galaxy which has an optical depth in stars that is of order unity (e.g. Kent \& Falco 1988, hereafter KF88; Schneider et al. 1988, hereafter S88; Schmidt, Webster \& Lewis 1998, hereafter SWL98). This results in a high probability for microlensing. 

Microlensing is indicated either by independent temporal variability of image fluxes, or by variation in colour between images at a single epoch (separated by the macro-image delay). Large, and rapid variation of the continuum flux is found in the variability record for Q2237+0305 (Irwin et al. 1989; Corrigan et al. 1991; $\O$stensen et al. 1996; Wozniak et al. 2000a,b). This variation has been used to argue that the optical emission region must be significantly smaller than the microlens Einstein Radius (ER), and therefore the typical scale of the caustic structure (e.g. Wambsganss, Paczynski \& Schneider 1990; Wyithe, Webster, Turner \& Mortlock 2000; Yonehara 2001). During a caustic crossing, a small source exhibits colour variability if the emission spectrum is scale dependent (e.g. Wambsganss \& Paczynski 1991; Fluke \& Webster 1999). Evidence of this effect from broad band observations of a microlensing event was presented by Corrigan et al. (1991). Furthermore, if the quasar emits in one wave-band at a scale much smaller than the ER, and on a scale larger than the ER in another band, then colour change may be seen in two random observations of a single image (particularly if the observations straddle a caustic crossing event), or between two different images, as a result of magnification of the smaller source. Ground based observations have confirmed differential amplification of the emission region. Lewis et al. (1998) determined the ratios of emission line equivalent widths relative to one image. They show $(i)$ the ratios remain fairly constant for one image from line to line, suggesting that the sizes of the emission regions for the lines are not greatly different, $(ii)$ that the ratios vary from image to image for a single epoch by a factor of $\sim 2.5$, and $(iii)$ that the ratio for a single image varies as a function of time, i.e. as a result of a microlensing event. These results are consistent with earlier results of Fillipenko (1989) who measured a $\sim25\%$ difference in the width of the MgII lines between the A and B images. 

The CIII] line, produced by extended broad line regions has been measured in an attempt to find the flux ratios $(R)$ using emission scales beyond the influence of microlensing (Yee \& De Robertis 1992; Racine 1992; Fitte \& Adam 1994; Saust 1994; Lewis et al. 1998). Mediavilla et al. (1998) observed the CIII] line in Q2237+0305 using two-dimensional spectroscopy, and found an arc (image of the extended narrow line region of the source) extending around three of the images, indicating a very extended region of emission. On the other hand, measurement of $R$ using the CIII] line is subject to differential extinction, and uncertainties in continuum subtraction. To avoid the effects of extinction, and possibly microlensing Falco et al. (1996) imaged Q2237+0305 at 3.6cm, finding flux ratios similar to those inferred from CIII]. In addition, Q2237+0305 has also been imaged in the Ultraviolet (Blanton, Turner \& Wambsganss 1998), and in X-rays (Wambsganss, Brunner, Schindler \& Falco 1999).

Many models have been proposed for the projected lens mass distribution based on observations of the lensed images of Q2237+0305 (e.g. KF88; S88; Kochanek 1991, hereafter K91; Wambsganss \& Paczynski 1994, hereafter WP94; Witt, Mao \& Schechter 1995, hereafter WMS95; SWL98; Chae, Turnshek \& Khersonsky 1998, hereafter CTK98). The majority of these predict a flux ratio for images B and A of $R_{BA}\sim 0.80 - 1.1$, consistent with the ratio measured in the radio of $R_{BA}=1.1\pm0.3$ (Falco et al. 1996). However, over the monitoring history, the optical light-curve shows variations in $R_{BA}$ between $\sim 0.2$ and $\sim 1.0$. The discrepancy is attributed primarily to microlensing, and firmly demonstrates that the optical flux ratios cannot be used as model constraints (e.g. S88; KF88; K91; WP94). Agol, Jones \& Blaes (2000) (hereafter AJB00) have found a mid-IR B:A flux ratio of $R_{BA}\sim1.1$. This ratio is consistent with observations in the radio (Falco et al. 1996). AJB00 interpreted their results as evidence for an extended region of mid-IR emission, with dimensions larger than the microlens Einstein Radius. 

In this paper we use microlensing models to calculate distributions of flux ratios for sources with different sizes and intensity profiles, and hence derive quantitative limits on the scale of the mid-IR emission. In Secs.~\ref{models} and \ref{macros} we describe the microlensing models, and summarise published macro-models for Q2237+0305. Sec.~\ref{results} discusses the methods used to infer the mid-IR source size from the observed optical and mid-IR flux ratios, and the source size limits implied by the published macromodels. Initially we restrict our discussion to the particular case of the flux ratio between images B and A. However we present results based on all image ratios in Sec.~\ref{sec_pairs}. In the conclusion we mention some implications for quasar physics.

\section{Microlensing models}
\label{models}

\begin{table}
\begin{center}
\caption{\label{simls}Values of the optical depth ($\kappa$) and the magnitude of the shear ($|\gamma|$)for each simulation. The number of stars ($N_*$), the theoretical mean magnification ($\langle \mu_th\rangle\equiv|(1-\kappa)^2-\gamma^2|^{-1}$), and the mean magnification of the map $(\langle \mu\rangle)$ are given in each case.}
\begin{tabular}{|l|c|c|c|c|c|}
\hline
model  & $\kappa$   & $|\gamma|$  &  $N_*$   &  $\langle \mu_{th} \rangle$  &  $\langle \mu \rangle$ \\ \hline
$i$   & 0.30      &  0.35       &  8374    &  2.72                        &  2.75                  \\
 $ii$     &  0.30      &  0.40       &  11022   &  3.03                        &  3.05              \\
$iii$      &  0.30      &  0.45       &  15471   &  3.48                        &  3.50                \\
$iv$      &  0.30      &  0.50       &  23724   &  4.17                        &  4.26               \\
$v$      &  0.35      &  0.35       &  13253   &  3.33                        &  3.48                 \\  
$vi$      &  0.35      &  0.40       &  18501   &  3.80                        &  3.95                 \\
$vii$      &  0.35      &  0.45       &  28262   &  4.55                        &  4.79                \\
$viii$      &  0.40      &  0.35       &  21709   &  4.21                        &  4.29               \\
$ix$      &  0.40      &  0.40       &  32993   &  5.00                        &  4.99                  \\
$x$      &  0.45      &  0.35       &  37964   &  5.55                        &  5.61                    \\
$xi$     &  0.69      &  0.71       &  16915   &  2.45                        &  2.40                  \\\hline

\end{tabular}
\end{center}
\end{table}

 Throughout the paper, standard notation for gravitational lensing is used. The Einstein Radius is defined to be the radius inside which the mean surface mass density is equal to the critical density. The Einstein Radius $\eta_{0}$ of a 1$M_{\odot} $ star projected into the source plane is
\begin{equation}
\eta_o = \sqrt{\frac{D_{ds}D_s}{D_d}\frac{4GM_{\odot}}{c^2}},
\end{equation}
where $G$ is the gravitational constant, $c$ is the speed of light, and $D_d$, $D_s$ and $D_{ds}$ are the angular diameter distances of the lens, source and from the lens to the source respectively. The normalised shear is denoted by $\gamma$, and the convergence or optical depth by $\kappa$. The model for gravitational microlensing consists of a very large sheet of point masses that simulates the section of the galaxy along the image line-of-sight, together with a shear term that includes the perturbing effect of the mass distribution of the lensing galaxy as a whole. The normalised lens equation for a field of point masses with an applied shear in terms of these quantities is
\begin{equation}
\vec{y}= \left( \begin{array}{cc}
	 1-\gamma & 0 \\
	0 & 1+\gamma 
	    \end{array} \right)\vec{x} + \sum_{j=1}^{N_{*}}m^{j}\frac{(\vec{x}^{j}-\vec{x})}{|\vec{x}^{j}-\vec{x}|^{2}}
\label{lens_map} 
\end{equation}
Here $\vec{x}$ and $\vec{y}$ are the normalised image and source positions respectively, and the $\vec{x}_{i}^{j}$ and $m^{j}$ are the normalised positions and masses of the microlenses. We have assumed that the entire optical depth is in compact objects, and that all microlenses have a common mass. The ray-tracing method (e.g. Kayser, Refsdal \& Stabell 1986; Schneider \& Weiss 1987; Wambsganss, Paczynski \& Katz 1990) was used to compute the magnification maps. Each map produced covered an area of $100\eta_{o}\times100\eta_{o}$, and was computed at a resolution of $2000\times2000$ pixels. The maps therefore describe the magnification distributions for a dynamic range in source size of $\sim 500$. 100 rays were traced per unlensed pixel. The region of the lens plane in which image solutions need to be found to ensure that $99\%$ of the total macro-image flux is recovered from a source point was described by Katz, Balbus \& Paczynski (1986). The union of areas in the lens plane corresponding to the flux collection area of each point in the area of the magnification is known as the shooting region, the method for determining the dimensions of which is described in Lewis \& Irwin (1995), and Wyithe \& Webster (1999). The radius of the disc of point masses was chosen to be 1.2 times that required to cover this shooting region.

Tab. \ref{simls} shows the parameters $\kappa$ and $\gamma$ for each of the 11 magnification maps produced, along with the number of stars used in the model, the theoretical mean magnification, and the mean magnification of the map. Models $i-x$ cover the $\kappa-\gamma$ parameter space near the values estimated for images A and B by SWL98. A model (corresponding to image C of SWL98) with larger $\kappa$ and $\gamma$ (model $xi$) is included for comparison, and as an indicator of the sensitivity of our results to the macro-parameters. 

\section{Macro-models for Q2237+0305}
\label{macros}

\begin{table*}
\begin{center}
\caption{\label{macromodels}Summary of published macro-models from the following authors Schneider et al. (1988) (S88) Kent \& Falco (1988) (KF88), Kochanek (1991) (K88), Rix, Schneider \& Bahcall (1992) (RSB92), Wambsganss \& Paczynski (1994) (WP94), Witt, Mao \& Schechter (1995) (WMS95), Schmidt, Webster \& Lewis (1998) (SWL98), Chae, Turnshek \& Khersonsky (1998) (CTK98). The labels corresponding to Figs.~\ref{limits_single}, \ref{limits} and \ref{pairs} are shown in column $ii$. Where a paper presents multiple models, they are labelled in column $iii$ using the authors nomenclature. The table presents values for the optical depth and shear (where available), the theoretical magnification for each image, and the magnification ratios.}

\begin{tiny}
\begin{tabular}{|c|c|c|c|c|c|c|c|c|c|c|c|c|c|c|c|c|c|c|c|}
\hline

      &      &        &   \multicolumn{3}{c}{A}  & \multicolumn{3}{c}{B}     &   \multicolumn{3}{c}{C}   & \multicolumn{3}{c}{D} & \multicolumn{3}{c}{R}  \\
Author&label&model   &$\kappa$&$|\gamma|$&$|\mu|$ & $\kappa$&$|\gamma|$&$|\mu|$ & $\kappa$&$|\gamma|$&$|\mu|$ & $\kappa$&$|\gamma|$&$|\mu|$ & $R_{BA}$         & $R_{AC}$         & $R_{DA}$   \\ \hline
S88   &S   &        &  0.36&0.44      & 4.6    & 0.45&0.28        & 4.5    &  0.88 & 0.53     & 3.8    & 0.61 & 0.66      & 3.6  &   0.98           &  1.20            &  0.78           \\
KF88  &KF  &De Vauc.&      &       & 3.34   &       &          & 2.64   &       &          &  1.14  &      &           &2.11 &0.88  &  2.63            &  0.63             \\
K91   &K1  &   1    &      &          & 10.0   &       &          & 6.8    &       &          & 7.20   &      &           & 8.10 &   0.68           &  1.43            &    0.81       \\
      &K2    &   2    &      &          & 4.3    &       &          &3.83    &       &          & 2.11   &      &           & 5.07 &   0.89             & 2.04             &  1.18           \\
      &K3    &   3    &      &          & 8.3     &      &          &7.30    &       &          & 4.65   &      &           & 8.88  &   0.88             & 1.78             &  1.07            \\
      &K4    &   4    &      &          & 2.3     &      &          & 2.05   &       &          & 1.08   &      &           & 2.23 &   0.89             & 2.13             &  0.97             \\
      &K5    &   5    &      &          & 2.6     &      &          & 1.72   &       &          & 1.66   &      &           & 1.72  &   0.66             & 1.56             &  0.66            \\
RSB92 &RSB1  &   1  &  0.42&0.42      & 5.9     &  0.41&0.41      & 6.1    &   0.71&0.63      & 3.1    &   0.65&0.60      & 4.1  &   1.03           & 1.92             &  0.69             \\
      &RSB2  &   2    &  0.39&0.46      & 5.8     &  0.38&0.46      & 6.3    &   0.65&0.67      & 3.0    &   0.59&0.67      & 3.6  &   1.09             & 1.96             &  0.62              \\
      &RSB2a &   2a   &  0.41&0.47      & 6.2     &  0.38&0.43      & 5.1    &   0.65&0.68      & 2.8    &   0.59&0.56      & 5.1  &   0.82             & 2.22             &  0.82             \\
      &RSB3  &   3    &  0.42&0.41      & 4.5     &  0.37&0.48      & 4.4    &   0.68&0.66      & 3.1    &   0.65&0.67      & 3.6  &   0.98             & 1.45             &   0.80             \\
WP94  &WP1 &$\beta=0.0$ &  0.00&0.77  & 2.43    &  0.00&0.73      & 2.12   &   0.00&1.36     & 1.16   &   0.00&1.19     & 2.38 &   0.88           & 2.08             &  0.98             \\
      &WP2   &$\beta=1.0$ &  0.47&0.41  &  8.98   &  0.46&0.40      & 7.71   &   0.56&0.63    & 5.04   &   0.51&0.58    & 9.64 &   0.86            & 1.79             &  1.07             \\
      &WP3   &$\beta=1.7$ &   0.84&0.13 &  103.6  &  0.84&0.13      & 87.80  &   0.88&0.17    & 66.0   &   0.86&0.17    & 118.5 &   0.85            & 1.56             &  1.14             \\
WMS95 &WMS1&$\beta=0.5$ &  0.23&0.60  &  4.3    &  0.22&0.60      & 3.8    &   0.29&0.96    & 2.4    &   0.26&0.88     & 4.4  &   0.88            & 1.79             &  1.03             \\
      &WMS2  &$\beta=1.0$ &  0.47&0.42  &   9.4   &  0.46&0.41      & 8.1    &   0.55&0.62    & 5.6    &   0.52&0.58    & 10.0 &   0.86            & 1.56             &  1.16             \\
      &WMS3  &$\beta=1.5$ &  0.73&0.22  &   36.5  &  0.72&0.21      &  31.3  &   0.79&0.30    & 23.3   &   0.76&0.29    & 40.7 &   0.86            & 1.56             &  1.16               \\
CTK98 &CTK1&$\epsilon=0.3$&     &      &   3.69  &     &           & 3.32   &        &         & 1.86   &        &         & 3.14  &   0.9           &  2.00            &     0.85           \\ 
      &CTK2 &$\epsilon=0.2$&     &      & 9.03    &     &           & 8.15   &        &         & 5.42   &        &         & 9.03 &   0.9           &  1.67            &     1.00           \\
      &CTK3 &$\epsilon=0.1$&     &      & 27.4    &     &           & 24.66  &        &         & 17.81  &        &         & 30.14 &   0.9             & 1.54           &    1.10          \\
SWL98 &SWL1&$\Lambda=0.5$& 0.36&0.40  & 3.90    & 0.36&0.42       & 4.39   &  0.69&0.71       & 2.42   & 0.59&0.61        & 4.91 &   1.12              &  1.63            &  1.26            \\
      &SWL2  &$\Lambda=1.0$&     &      & 4.05    &     &           & 4.37   &        &         & 2.47   &        &         & 5.10 &   1.08              &  1.64            &  1.26             \\
      &SWL3  &$\Lambda=2.0$&     &      & 4.18    &     &           & 4.34   &        &         & 4.35   &        &         & 2.51 &   1.04             &  1.67            &  1.24              \\\hline

\end{tabular}
\end{tiny}

\end{center}
\end{table*}

The quadruple quasar Q2237+0305 has been the target of numerous studies. The authors have attempted to explain the geometry of the system via gravitational lensing using various models. S88 and later RSB92 assumed that mass in the lensing galaxy was proportional to observed light. Other approaches parameterise the lens using various simple potentials having either intrinsic or external quadrupoles (KF88; K91; WP94; WMS95; SWL98; CTK98). Constraints on the lens model come from the position of the galaxy and of the 4 quasar images. In addition, the flux ratios potentially provide 3 additional constraints. However, these are difficult to measure from optical images due principally to the effects of differential absorption, and microlensing. KF88 used the available optical flux ratios, while noting that the mismatch of the flux for images C and D is the dominant contribution to the error in their fit. Optical flux ratios were also used by S88 and RSB92, though the minimisation quantity was weighted 9:1 in favour of image positions. More reliable flux ratios, from the CIII] line (Racine 1992) were employed by WMS95, and more recently, CTK98 have incorporated the 3.6cm fluxes measured by Falco et al. (1996). Other studies fitted only image positions. All these studies have reproduced the best measured positions (currently those of Blanton et al. 1998) of the time to within (or near to) the quoted error. However flux ratios, particularly in the optical have been less successful. A major conclusion of the parametric studies is that the image configuration (and flux ratios) can be reproduced by families of models, and that the image geometry alone is insufficient to uniquely compute the optical depth and shear at the location of each image. Hence parametric lens models do not uniquely define the individual image magnifications nor the total magnification. Indeed, different potentials predict total magnifications ranging from a few to a few hundred (WP94; WMS95; CTK98). On the other hand, the total mass inside the Einstein ring is consistently measured at $\sim1.5\times10^{10}M_{\odot}h^{-1}_{75}$ (RSB92; WP94; CTK98). In addition, the large variation in total magnification is not found in the magnification ratios. Most published models have flux ratios for images B and A of $\sim0.80-1.1$, and the degenerate model families of WP94, WMS95 and CTK98 all have $R_{BA}\sim0.85-0.90$. In a more physically motivated parametric model, SWL98 included a bar component in the lensing galaxy to correct for the apparent misalignment between the model major axis and observed galactic orientation. SWL98 broke the degeneracy in their family of models (analogous to the shear - power-law degeneracy discussed by WP94) through the use of an elliptical mass model having the observed eccentricity of the light distribution. The collection of models from the literature described in this section are summarised in Tab.~\ref{macromodels}.

\section{mid-IR source size limits from flux ratios}
\label{results}

\begin{figure*}
\vspace*{195mm}
\includegraphics{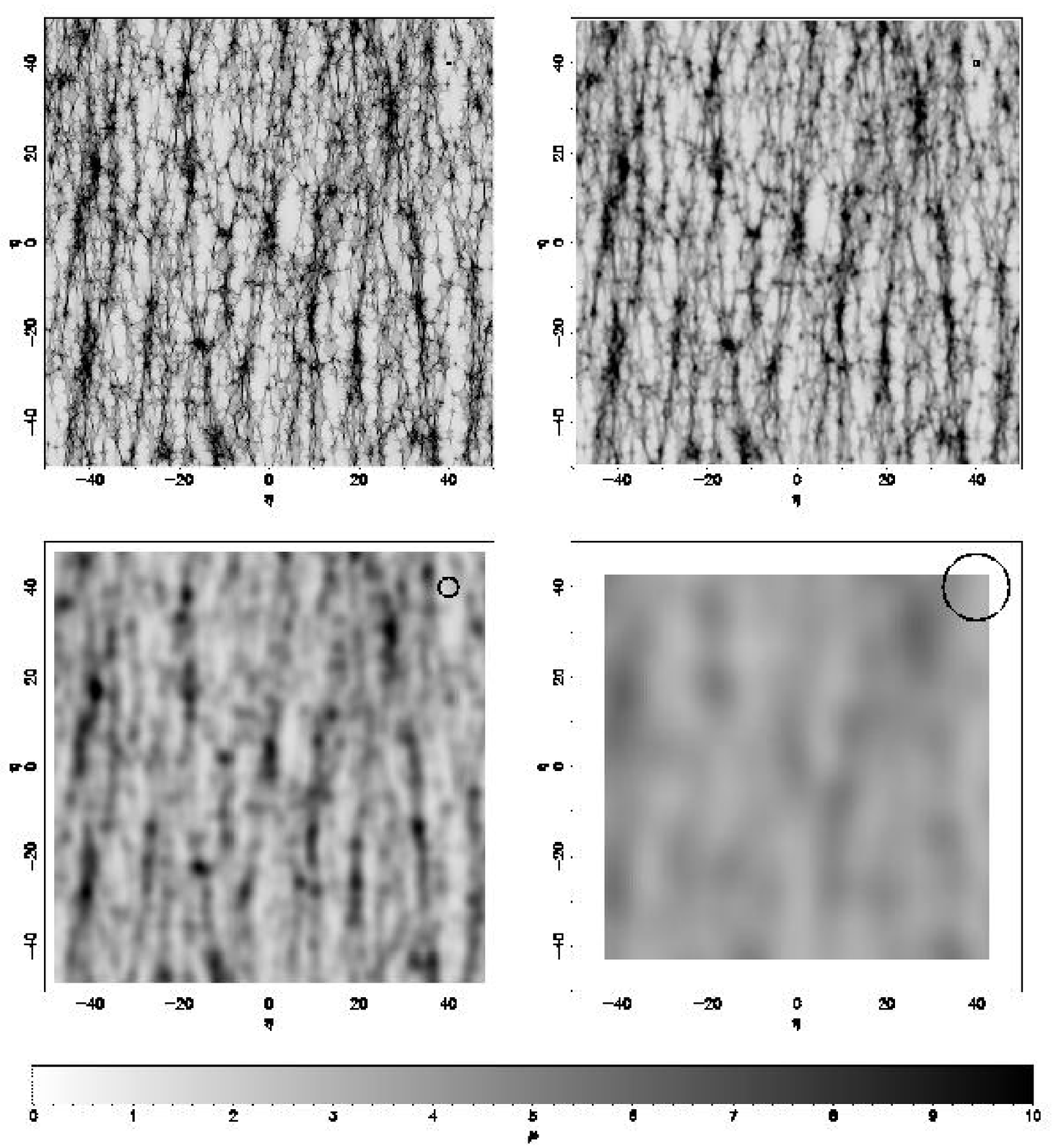}
\caption{\label{magmap} Convolution of a magnification map with source discs of 4 different sizes (shown by the circles in the upper right corner of each panel, the radii are $S$=0.15, 0.60, 2.10 and 7.35$\eta_o$). The grey-scale bar shows the corresponding magnification. The macrolensing parameters were $\kappa=0.35$, $\gamma=0.40$. The axes are in units of $\eta_o$.}
\end{figure*}

\begin{figure}
\vspace*{65mm}
\includegraphics{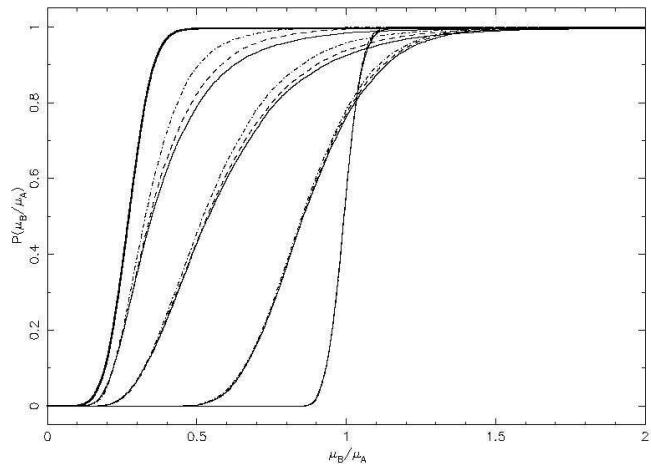}
\caption{The cumulative probability of the mid-IR flux ratio $R_{BA}^{IR}=\mu_B/\mu_A$ as a function of mid-IR source size ($S_{IR}$) for a variety of Gaussian source sizes $S_{IR} = 2\sigma = 0.4, 1.6, 6.4, 25.6\eta_o$ (left to right). The parameters for both images A and B were $\kappa=0.35$, $\gamma=0.40$. Plots are shown for a uniform optical source: $S_{OPT} = 0.025\eta_o$ (single pixel, solid lines), and 2 Gaussian sources: $S_{OPT}=2\sigma = 0.05\eta_o$ (dashed lines) and $S_{OPT}=2\sigma = 0.1\eta_o$ (dot-dashed lines). The thick line represents the prior for optical flux ratio.}
\label{probfns} 
\end{figure}

\begin{figure*}
\vspace*{65mm}
\includegraphics{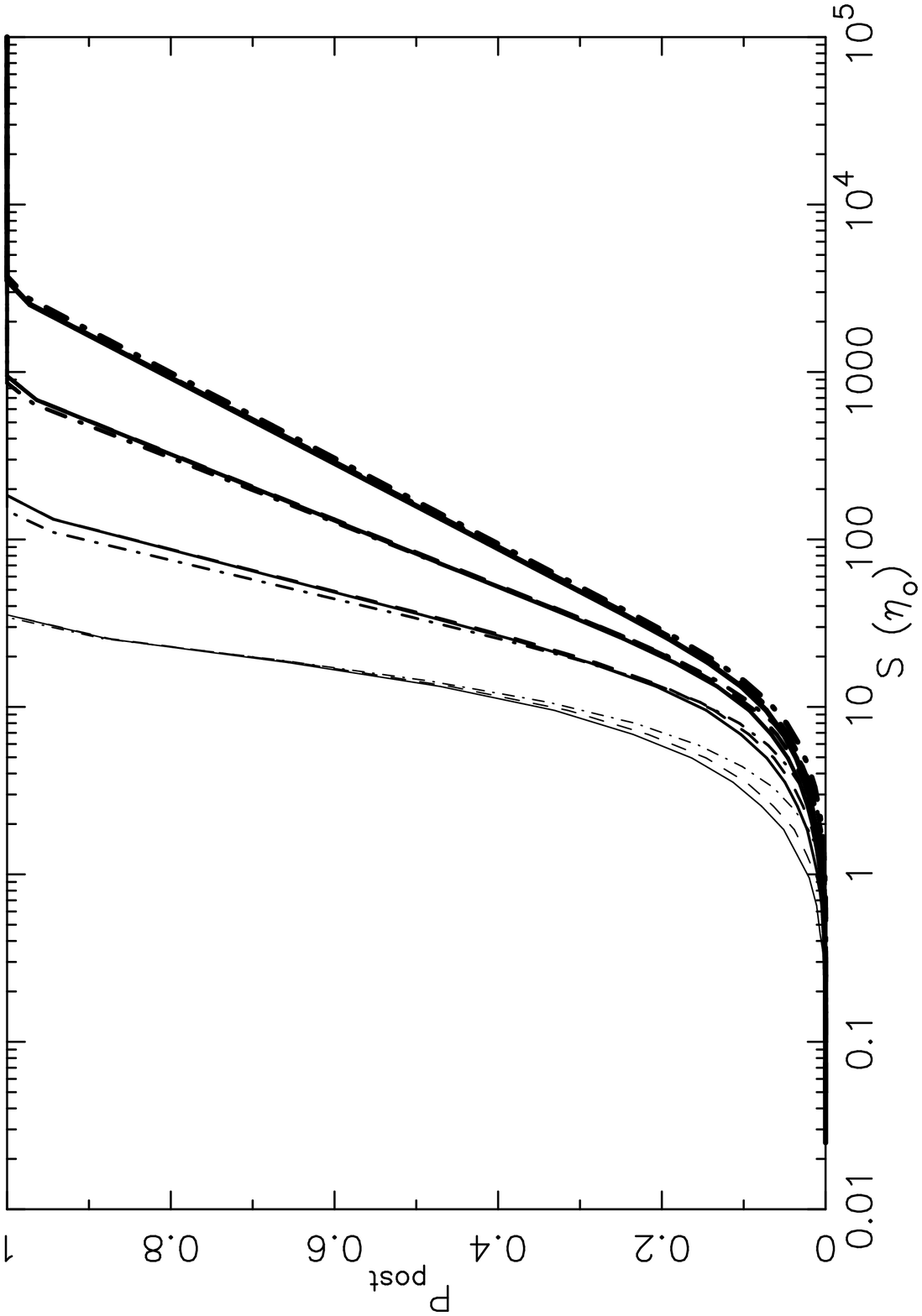}
\includegraphics{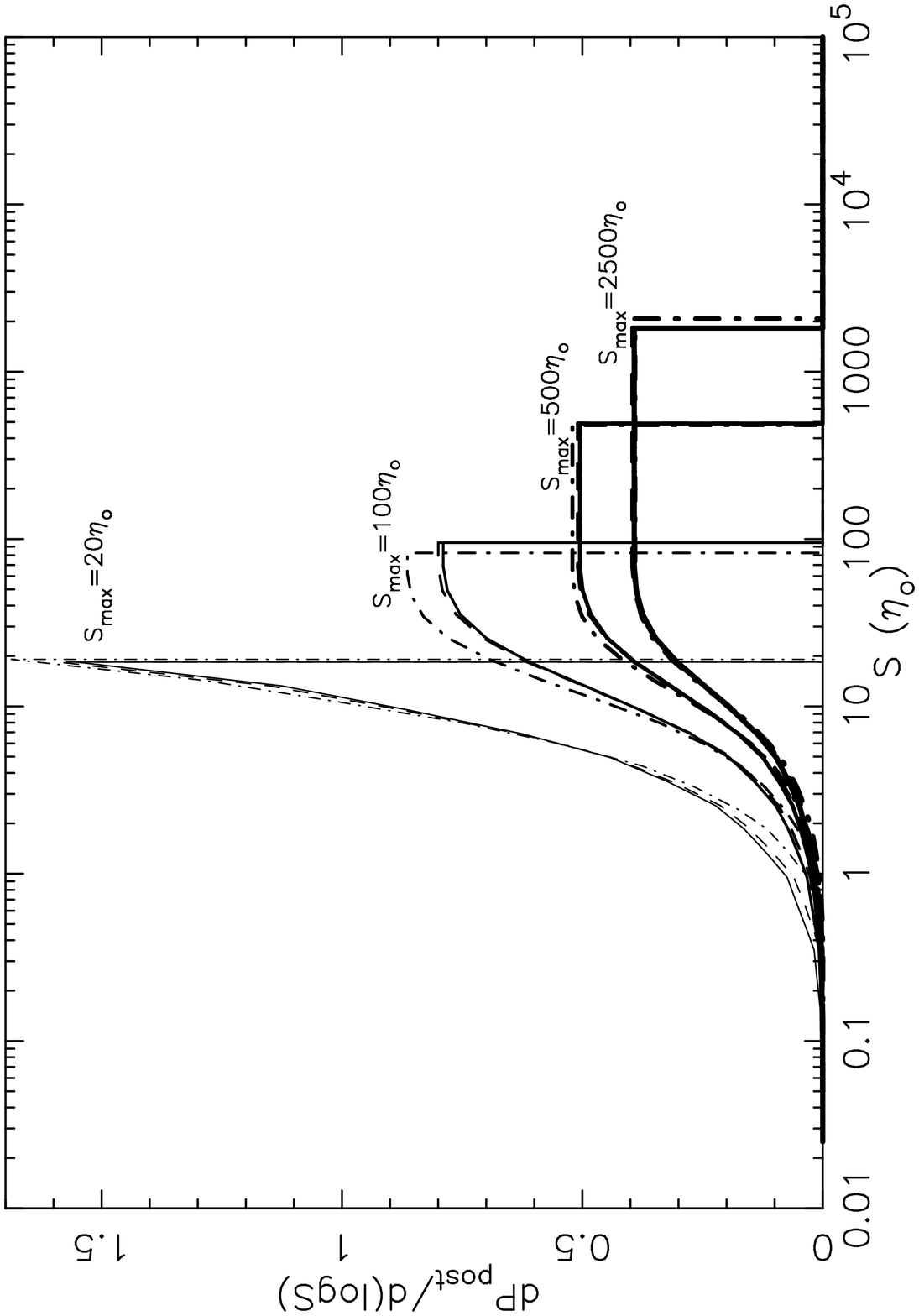}
\caption{Left: The cumulative probability for the IR source size $P\left(S_{IR}<S|S_{OPT}\right)$, and Right: The differential probability for $S_{IR}$ given a uniform optical source: $S_{OPT}= 0.025\eta_o$ (single pixel, solid lines), and 2 Gaussian sources:  $S_{OPT}=2\sigma = 0.05\eta_o$ (dashed lines) and $S_{OPT}=2\sigma = 0.1\eta_o$ (dot-dashed lines). The parameters for both images A and B were $\kappa=0.35$, $\gamma=0.40$. The mid-IR source was also assumed to be Gaussian with halfwidth $\sigma=S_{IR}/2$.}
\label{probfns2} 
\end{figure*}

\begin{figure*}
\vspace*{135mm}
\includegraphics{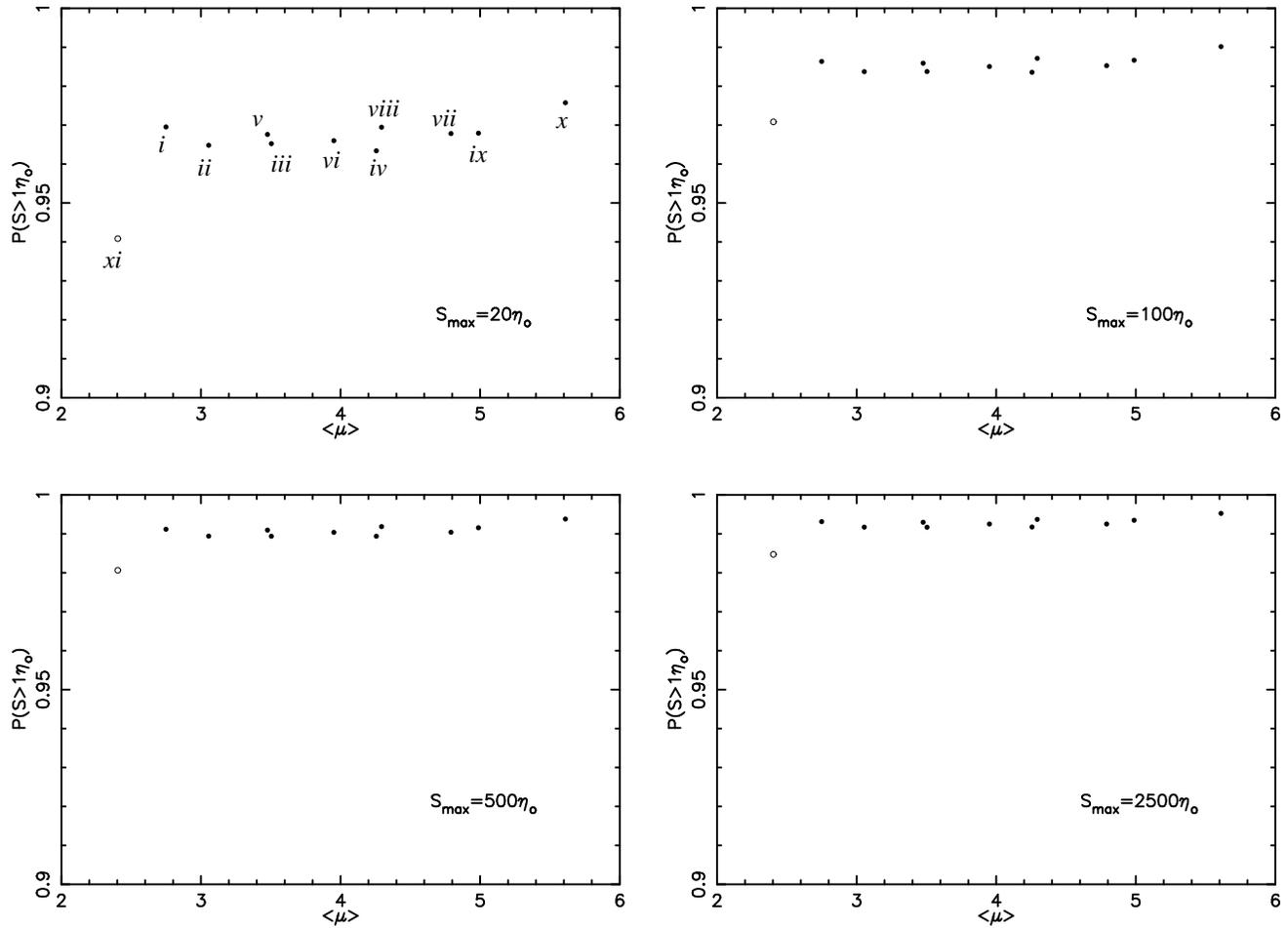}
\caption{The probability that the IR source size is greater than $1\eta_o$ ($1-P\left(S_{IR}<1\eta_o|S_{OPT}\right)$) as a function of mean magnification. The parameters for both images A and B were the same in each case, the macro parameters were from models $i-x$ (dark dots) and model $xi$ (open circle). The optical source was uniform with $S_{OPT}= 0.025\eta_o$ (single pixel). The mid-IR source was also assumed to be Gaussian with halfwidth $\sigma=S_{IR}/2$.}
\label{probfns4} 
\end{figure*}

\begin{figure*}
\vspace*{100mm}\includegraphics{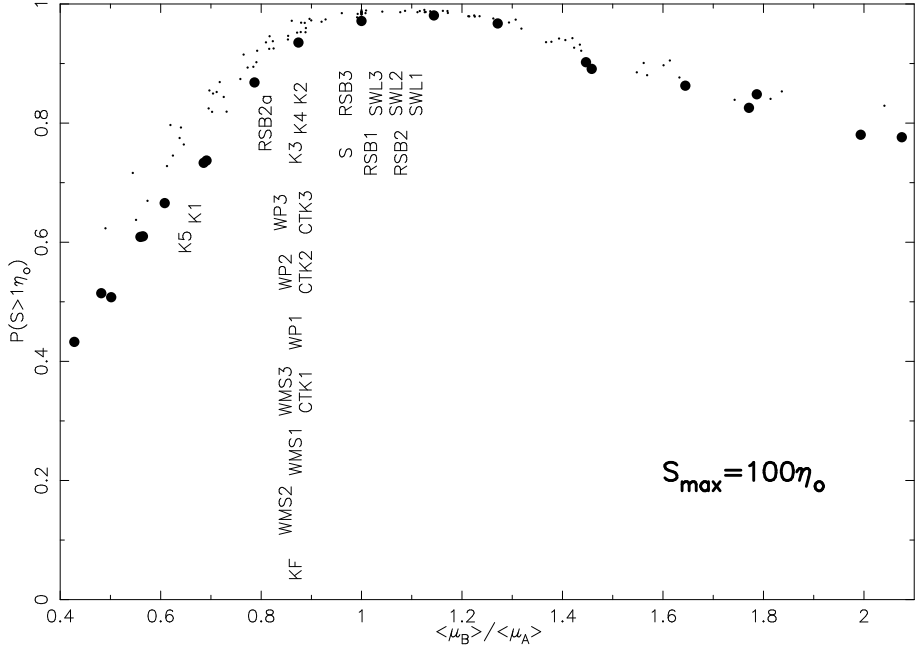}
\caption{\label{limits_single} Plot of the probability that $S_{IR}>1.0\eta_o$ as a function of mean flux ratio, $S_{OPT}=0.025\eta_o$ (single pixel). The larger markers indicate models that include the large $\kappa$ large $\gamma$ for one of the images (model $xi$). The labels correspond to published macro-models (see Tab.~\ref{macromodels}) and are plotted at the appropriate $R_{AB}$.}
\end{figure*}

\begin{figure*}
\vspace*{185mm}
\includegraphics{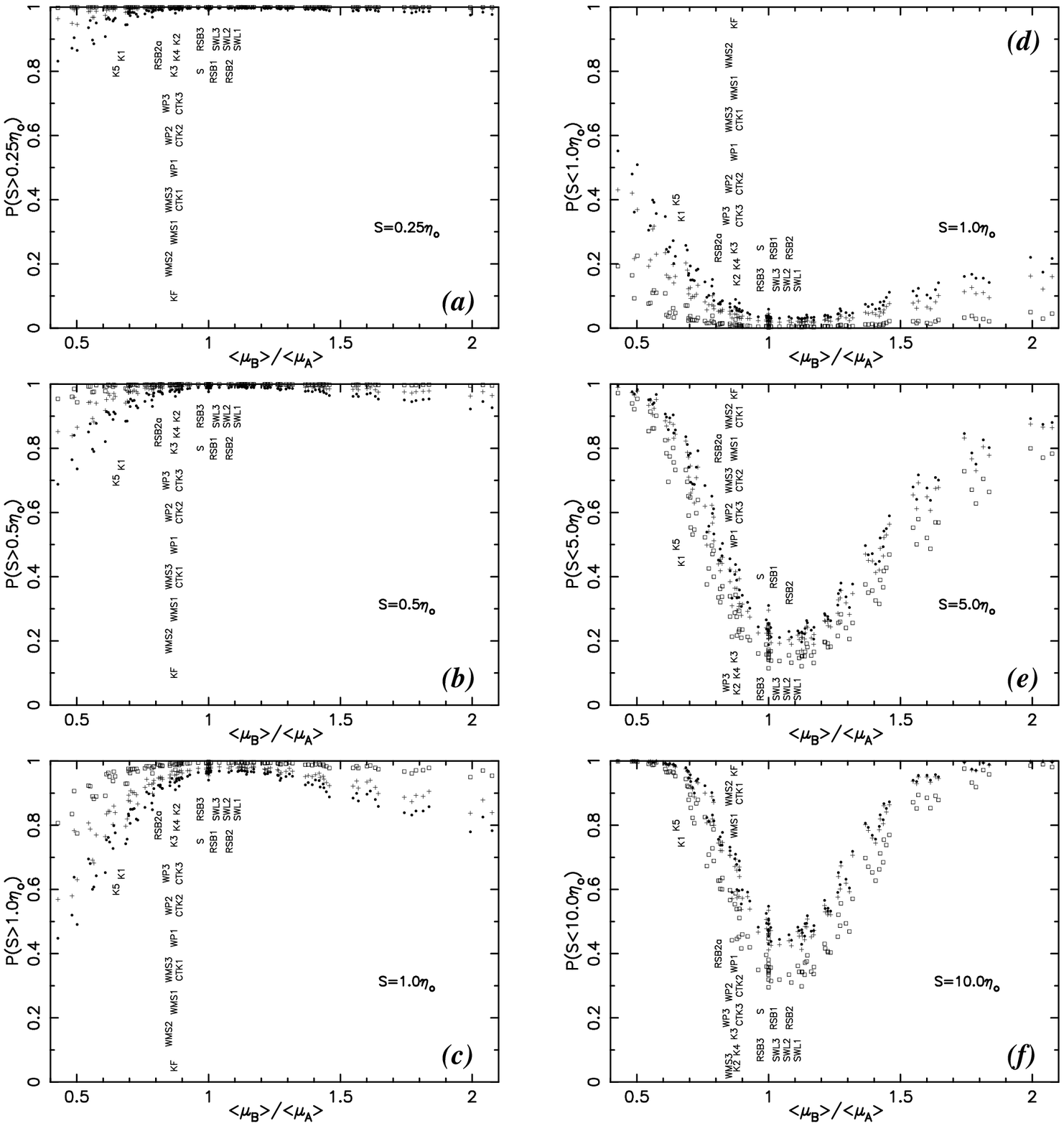}
\caption{\label{limits} Plots of source size limits as a function of mean flux ratio. Results for three assumptions of optical size are shown; A uniform source: $S_{OPT}=0.025\eta_o$ (single pixel, dots), and two Gaussian sources: $S_{OPT}=2\sigma=0.05\eta_o$ (crosses) and $S_{OPT}=2\sigma=0.1\eta_o$ (squares). Panels (a)-(c): Lower limits ($S_{IR}>$0.25, 0.5 and 1.0$\eta_o$). Panels (d)-(f): Upper limits ($S_{IR}<$1.0, 5.0 and 10.0$\eta_o$). The larger markers indicate models that include the large $\kappa$ large $\gamma$ for one of the images (model $xi$). Note that panel (d) = one minus panel (c).}
\end{figure*}

\begin{table*}
\begin{center}
\caption{\label{flux_fractions} Values for the fractions of total flux in each image, and the flux ratios in the mid-IR (AJB00) and V-bands (Wozniak et al. 2000a,2000b). The V-band fluxes have been de-reddened (AJB00), and the dominant contribution to the quoted error comes from this procedure.}
\begin{tabular}{|c|c|c|c|c|c|}
\hline
                   &           & \multicolumn{4}{c}{Fraction of Flux}  \\
Wavelength         & Date (1999)     & A           & B           & C           & D           \\\hline
8.9 \& 11.7$\mu m$ & 28th July \& 21st September & 0.27$\pm$.02& 0.30$\pm$.02& 0.16$\pm$.02& 0.27$\pm$.02\\
V-band             & 1st August       & 0.39$\pm$.08& 0.11$\pm$.02& 0.41$\pm$.14& 0.10$\pm$.03 \\ 
V-band             & 26th September      & 0.46$\pm$.10& 0.12$\pm$.02& 0.32$\pm$.11& 0.10$\pm$.03\\\hline
\end{tabular}
\end{center}

\begin{center}
\begin{tabular}{|c|c|c|c|c|c|c|c|}
\hline
                   &           & \multicolumn{6}{c}{Flux Ratio} \\
Wavelength         & Date (1999)     & $R_{BA}^{obs}$         & $R_{AC}^{obs}$         & $R_{DA}^{obs}$         & $R_{BC}^{obs}$         & $R_{DB}^{obs}$         & $R_{DC}^{obs}$         \\\hline
8.9 \& 11.7$\mu m$ & 28th July \& 21st September & 1.1$\pm$.1& 1.7$\pm$.3& 1.0$\pm$.1& 1.9$\pm$.3& 0.9$\pm$.1& 1.7$\pm$.2 \\
V-band             & 1st August     & 0.28$\pm$.08& 0.95$\pm$.37& 0.26$\pm$.09& 0.27$\pm$.09& 0.91$\pm$.31& 0.24$\pm$.10 \\ 
V-band             & 26th September & 0.26$\pm$.07& 1.43$\pm$.55& 0.22$\pm$.08& 0.38$\pm$.15& 0.83$\pm$.29& 0.31$\pm$.13 \\   \hline            
\end{tabular}
\end{center}
\end{table*}

The magnification of an extended microlensed source is calculated by convolving the source intensity profile with a magnification map. Fig.~\ref{magmap} shows a magnification map ($\kappa=0.35,$ $\gamma=0.40$) convolved with discs having radii of $S$=0.15, 0.60, 2.10 and 7.35$\eta_o$. Circles corresponding to these sizes are plotted in the top-right corner of each map. The grey-scale bar describes the magnification. Fig.~\ref{magmap} demonstrates that even a source several ER in extent can be microlensed by an appreciable factor (see for example Refsdal \& Stabell 1993). In addition, there is a correlation between the caustic regions of highest density and the higher magnification regions for the larger sources. However the figures demonstrate that there is significant loss of correlation between the magnifications of a source which is $\ll\eta_o$ and of larger co-positional sources. In particular, larger sources show light-curves with lower event amplitudes and longer event durations (see the light-curves and correlation functions for various source sizes in Wambsganss, Paczynski \& Katz 1990).

The cumulative probability for the mid-IR flux ratio $R_{BA}^{IR}=\frac{\mu_B^{IR}}{\mu_A^{IR}}$, for different mid-IR source sizes $S_{IR}$ given the measured optical ratio is
\begin{eqnarray}
\label{ratioprob}
\nonumber
&&\hspace{-7mm}P_R(R_{BA}^{IR}<\frac{\mu_B}{\mu_A}|S_{IR},S_{OPT})=\int_0^{\infty}dR_{BA}^{OPT,obs}p(R_{BA}^{OPT,obs})\\
&&\hspace{-7mm}\times P_R(R_{BA}^{IR}<\frac{\mu_B}{\mu_A}|S_{IR},S_{OPT},R_{BA}^{OPT,obs}).
\end{eqnarray}
$p(R_{BA}^{OPT,obs})$ is the probability for the observed optical flux ratios, which we have assumed to be Gaussian with a mean and halfwidth ($\sigma$) equal to the observed value and uncertainty. The integral over $R_{BA}^{OPT,obs}$ was performed via Monte-Carlo. The observed optical flux fractions and ratios are summarised in Tab.~\ref{flux_fractions}. In the V-band (Wozniak et al. 2000b), the de-reddened flux ratios (AJB00) were $R_{BA}^{OPT,obs}=0.28\pm0.08$ on the 1st August 1999 and $0.26\pm0.07$ on the 26th September 1999. In the remainder of this paper we use the V-band fluxes from the 26th of September 1999. Fig.~\ref{probfns} shows the cumulative probability for mid-IR flux ratio $P_R(R_{BA}^{IR}<R|S_{IR},S_{OPT})$ for four different mid-IR source sizes $S_{IR}$ given three different optical source sizes $S_{OPT}$ in each case. The mid-IR sources were assumed to be Gaussian with half width $\sigma$: $S_{IR}=2\sigma=0.4,1.6,6.4, 25.6\eta_o$ (left to right). The optical sources were uniform: $S_{OPT}=0.025\eta_o$ (single pixel), and Gaussian: $S_{OPT}=2\sigma=0.05\eta_o$ (dashed lines) and $S_{OPT}=2\sigma=0.1\eta_o$ (dot-dashed lines). The parameters for both images A and B were $\kappa=0.35$, $\gamma=0.40$. The thick line shows $p(R_{BA}^{OPT,obs})$. Small mid-IR sources have distributions of $R_{BA}^{IR}$ with means close to the observed optical ratio, while larger sources have distributions which are narrow, and symmetric about the macro-model flux ratio $R_{BA}$ (1 for the example in Fig.~\ref{probfns}). The symmetry signifies a loss of correlation between the flux ratio of the large mid-IR source and the smaller V-band source. Mid-IR source sizes of $\sim1\eta_o$ show intermediate behaviour. $R_{BA}^{IR}$ for a given $S_{IR}$ is typically larger for smaller assumed $S_{OPT}$ due to the lower correlation of mid-IR and optical fluxes.

Source size limits are imposed by the difference in the flux ratios for the optical source (which is known to be $\ll\eta_o$) and the mid-IR source. Given the observed V-band flux ratio ($R_{BA}^{OPT,obs}$), the probability that the mid-IR flux ratio ($R_{BA}^{IR}$) of two images B and A is smaller than the observed value $R_{BA}^{IR,obs}$ is not a monotonic function of the mid-IR source size $S_{IR}$ for $R_{BA}^{IR}$ greater than $R_{BA}$ (the macro-model flux ratio). However, the $R_{BA}^{IR}$ for sources with $S_{IR}\sim\eta_o$ are quite uncorrelated with $R_{BA}^{OPT}$.

For each mid-IR source size we construct the likelihood 
\begin{eqnarray}
\label{lhood}
\nonumber
&&\frac{dP_{lh}}{dR_{BA}^{IR}}(R_{BA}^{IR}|S_{IR},S_{OPT}) = \\
&&N\int_0^{\infty}dR_{BA}^{IR,obs}  \frac{dP_R}{dR}(R_{BA}^{IR}=R_{BA}^{IR,obs})p(R_{BA}^{IR,obs}).
\end{eqnarray}
$p(R_{BA}^{IR,obs})$ is the probability for the observed mid-IR flux ratio, which we have assumed to be Gaussian with a mean and halfwidth ($\sigma$) equal to the observed value and uncertainty, and $\frac{dP_R}{dR}$ is from Eqn.~\ref{ratioprob}. The observed flux fractions and ratios are summarised in Tab.~\ref{flux_fractions}. In the mid-IR (both 8.9 and 11.7 $\mu m$), AJB00 quote $R_{BA}^{IR,obs}=1.1\pm0.1$ for observations close to the V-band observations of Wozniak et al. (2000) (28th July and 21st September). The infinity in the integral over $R_{BA}^{IR,obs}$ was approximated by a value $>2R_{BA}^{IR,obs}$. Eqn.~\ref{lhood} is a function of the optical source size ($S_{OPT}$) assumed. $S_{OPT}$ is thought to be $\la10^{-2}\eta_o$ from a variety of arguments (e.g. Wambsganss, Paczynski \& Schneider 1990; Wyithe, Webster, Turner \& Mortlock 2000). Rather than convolve with a further probability for $S_{OPT}$, we have computed the likelihood (Eqn.~\ref{lhood}) for several values of $S_{OPT}<0.1\eta_o$. 

To construct probabilities for source size we have combined Eqn.~\ref{lhood} with uniform logarithmic Bayesian priors
\begin{eqnarray}
\label{prior}
\nonumber
\frac{dP_{prior}}{dlog(S_{IR})}&\propto& 1  \hspace{5mm} {\rmn where} \hspace{5mm} S_{IR}<S_{max}\\
                               &=& 0 \hspace{5mm} {\rmn otherwise}
\end{eqnarray}
for the unknown mid-IR source size $S_{IR}$:
\begin{eqnarray}
\label{size_lim}
\nonumber 
&&P_{S}(S<S_{IR}|S_{OPT}) = \\
&&N\int_0^{S_{IR}}dS'\frac{dP_{prior}}{dS}  \frac{dP_{lh}}{dR_{BA}^{IR}}(R_{BA}^{IR}|S',S_{OPT}).
\end{eqnarray}
$N$ is a normalising constant. Fig.~\ref{probfns2} shows the cumulative distribution $P_S(S_{IR}<S|S_{OPT})$ as well as the differential probability for source size corresponding to the example in Fig.~\ref{probfns}. The distributions are plotted assuming 4 different upper cutoffs in the prior for source size, $S_{max}=20, 100, 500$ and $2500\eta_o$. $S_{max}$ is known to be smaller than $\sim 25000\eta_o\sqrt{\frac{M_{\odot}}{\langle m\rangle}}$ since it is not resolved in the imaging data (AJB00). For this macro model, the source is larger than $S_{IR}=1\eta_o$ with greater than 97\% ($S_{OPT}=0.025\eta_o$) confidence. The probability shows minimal dependence on $S_{OPT}$. In this case the measured $R_{BA}^{IR,obs}$ differs from $R_{BA}$ only by the observational uncertainty, and therefore carries no information on how large the source can be. As a result the upper limits are entirely dependent on the upper cutoff for the prior assumed. The upper cutoff is the parameter to which the probabilities are most sensitive. Cumulative distributions $P_S(S_{IR}<S|S_{OPT})$ were computed where images A and B are assumed to have identical parameters, hence $R_{BA}=1$ in each case. The confidence that $S_{IR}>1\eta_o$ is plotted as a function of the macro-model magnification in Fig.~\ref{probfns4}, assuming 4 different cutoffs $S_{max}$, the open circle denotes model $xi$. There is some spread in these values, (e.g. $P_S\sim 0.94-0.97$ for $S_{max}=20\eta_o$ and $P_S\sim 0.98-0.99$ for $S_{max}=500\eta_o$), and higher mean model magnification tends to result in slightly tighter constraints. 

$P_S(S<S_{IR}|S_{OPT})$ was computed for all 121 combinations of the macro-parameters $i-xi$. Fig.~\ref{limits_single} shows the confidence that $S_{IR}>1\eta_o$ as a function of the macro-model flux ratio. Probabilities corresponding to $S_{max}=100\eta_o$ are plotted (similar results are obtained for other values of $S_{max}$. Different assumptions for macro-models provide different limits since large sources have a flux ratio that is correlated more with the macro-model flux-ratio, and less with the optical flux ratio. Fig.~\ref{limits_single} demonstrates a correlation between probability and $R_{BA}$ that has surprisingly little scatter. When considering source size limits determined from flux ratios, we can therefore approximately parameterise different macro models by their predicted flux ratio. The larger markers in Fig.~\ref{limits_single} denote those probabilities computed assuming one or both image macro-models to be model $xi$. These lie below the other points, illustrating the generalisation of the correlation in Fig.~\ref{probfns4}. However, because the limits obtained from macro-models that include model $xi$ follow nearly the same correlation as combinations of models $i-x$, the correlation may be valid for combinations of $\kappa_A/\gamma_A$ and $\kappa_B/\gamma_B$ other than those investigated. 

Distributions of flux ratios vary with macro-model for fixed $R_{BA}$. Therefore the correlation in Fig.~\ref{limits_single} is presumably due to the imposition of the optical flux ratio. The following approximation demonstrates why this should be the case. Consider two microlensed images $b$ and $a$. Suppose that the magnification map for image $a$ is uniform, while the magnification map for image $b$ contains a single caustic but is otherwise uniform. Let the magnification of image $a$ be $\mu^o_{a}$. Furthermore, let the magnification of image $b$ be comprised of the magnification due to critical images associated with the caustic $\mu^c_b$ in addition to the magnification from non-critical images $\mu^o_b$. The resulting optical magnification ratio is 
\begin{equation}
R_{ba}^{OPT} = \frac{\mu_b^{OPT,o} + \mu_b^{OPT,c}}{\mu_{a}^{OPT,o}} \sim R_{ba} +  \frac{\mu_b^{OPT,c}}{\mu_{a}^{OPT,o}},
\end{equation}
where $R_{ba}$ is the theoretical flux ratio. Similarly, the larger mid-IR source size has a magnification ratio
\begin{equation}
R_{ba}^{IR} \sim R_{ba} +  \frac{\mu_b^{IR,c}}{\mu_{a}^{IR,o}}.
\end{equation}
Now $\mu_{a}^{OPT,o}=\mu_{a}^{IR,o}\equiv\mu_{a}^o$ by construction, so $R_{ba}^{IR} \rightarrow R_{ba}$ for large mid-IR sources. Consider a different model having both an increased $\mu^o_{a}$ and $\mu_{b}^{OPT,o}$ while keeping $R_{ba}$ constant. To maintain the observed $R_{ba}^{OPT}$, $\mu_b^{OPT,c}$ must be increased, either by increasing the caustic strength (flux factor, see Witt (1990)), or by moving the optical source closer to the caustic. In the former case $\mu_b^{IR,c}$ is increased by the same fraction as $\mu_b^{OPT,c}$, thus the same source size is required to produce the observed flux ratio for different models having the same $R_{ba}$. In the second case, shifting the larger mid-IR source relative to the caustic has a smaller effect on $\mu_b^{IR,c}$ than on $\mu_b^{OPT,c}$, thus increasing $\mu_b^{OPT,o}$ and $\mu_{a}^{OPT,o}$ slightly lowers the resulting $R_{ba}^{IR}$. A larger mid-IR source is therefore required to reproduce the observed mid-IR flux ratio, and hence the confidence for each (upper) limit is increased. The combination of these two effects suggests an explanation for both the trend (of tighter constraints with higher mean model magnification) shown in Fig.~\ref{probfns4}, and the correlation seen in Fig.~\ref{limits_single}.

The left-hand panels of Fig.~\ref{limits} show the probabilities (lower limits) that $S_{IR}>0.25\eta_o$, $S_{IR}>0.5\eta_o$ and $S_{IR}>1.0\eta_o$ as a function of the macro-model flux ratio $R_{BA}$. Similarly, the right-hand 3 panels of Fig.~\ref{limits} show the confidences (upper limits) that $S_{IR}<1.0\eta_o$, $S_{IR}<5.0\eta_o$ and $S_{IR}<10.0\eta_o$. Points are shown representing the three assumptions for the optical source size, a uniform source $S_{OPT}=0.025\eta_o$ (one pixel, dots), and 2 Gaussian sources $S_{OPT}=2\sigma=0.05\eta_o$ (crosses) and $S_{OPT}=2\sigma=0.1\eta_o$ (squares). The confidences show some dependence on $S_{OPT}$. Stronger limits are obtained if the optical source is larger since it is then not subject to such large fluctuations. In the remainder of this paper, we quote limits based on the assumption of $S_{OPT}=0.025\eta_o$.
 
\subsection{IR-source size limits} 

The statistics of flux ratios, particularly for large sources are sensitive to the macro-model assumed. Rather than select a particular model, or attempt to assign relative weights to limits found using different published models, we have computed the statistics for an ensemble of microlensing models for the images A and B (models $i-xi$ in Tab.~\ref{simls}). These models cover the $\kappa-\gamma$ parameter space near the values estimated for images A and B by SWL98 (models $i-x$), as well as one with larger $\kappa$ and $\gamma$ (corresponding to image C of SWL98, model $xi$). The models have mean magnifications ranging from 2.72-5.56. Combinations of these models for images A and B of Q2237+0305 yield macro-model flux ratios $R_{BA}$ that differ by a factor of $>2$ in both directions from the value observed in the mid-IR by AJB00. Most of the published macro-models (Tab.~\ref{macromodels}) lie within the covered region, both in terms of values for $\kappa$ and $\gamma$, and $\mu_{tot}$. The exceptions are those models in the degenerate families, with very large magnifications (WP94; WMS95) resulting from a profile which is flatter than isothermal. We note tentatively that Figs.~\ref{probfns4} to \ref{limits} show a weak correlation of confidence with magnification. These high magnification models will therefore yield higher confidences due to the denser caustic network. The models of WP94 and WMS95, which have profiles significantly steeper than isothermal produce a total magnification within the range of our ensemble of macro-models, but with smaller $\kappa$ and larger $\gamma$.

The predicted $R_{BA}$ of the macro models (Tab.~\ref{macromodels}) are marked on Figs.~\ref{limits_single} and ~\ref{limits}. Most models predict an average flux ratio of between 0.8 and 1.1. From Fig.~\ref{limits}, this implies that, for a uniform optical source with $S_{OPT}=0.025\eta_o$, the mid-IR source has $S_{IR}>1\eta_o$ with $>90\%$ confidence, and $>0.5\eta_o$ with $>95\%$ confidence. The model of SWL98 yields an average flux ratio of $\sim 1.0-1.1$, and hence an estimate of $S_{IR}>1\eta_o$ with $>97\%$ confidence. A Gaussian source with a radius $S_{IR}=2\sigma$ twice that of the assumed optical region is inconsistent with the results at $>99\%$ for all but 2 models (K1 and K5 for which we find limits of $>$98\%). The mid-IR and optical flux may therefore be said to come from different emission scales with high confidence. The confidences on upper limits are far more systematically dependent on the theoretical flux ratio, ranging between 80\% and 15\% for the $5.0\eta_o$ source, and 95\% and 30\% for the $10.0\eta_o$ source.

\subsection{Different source profiles} 

\begin{figure*}
\vspace*{185mm}
\includegraphics{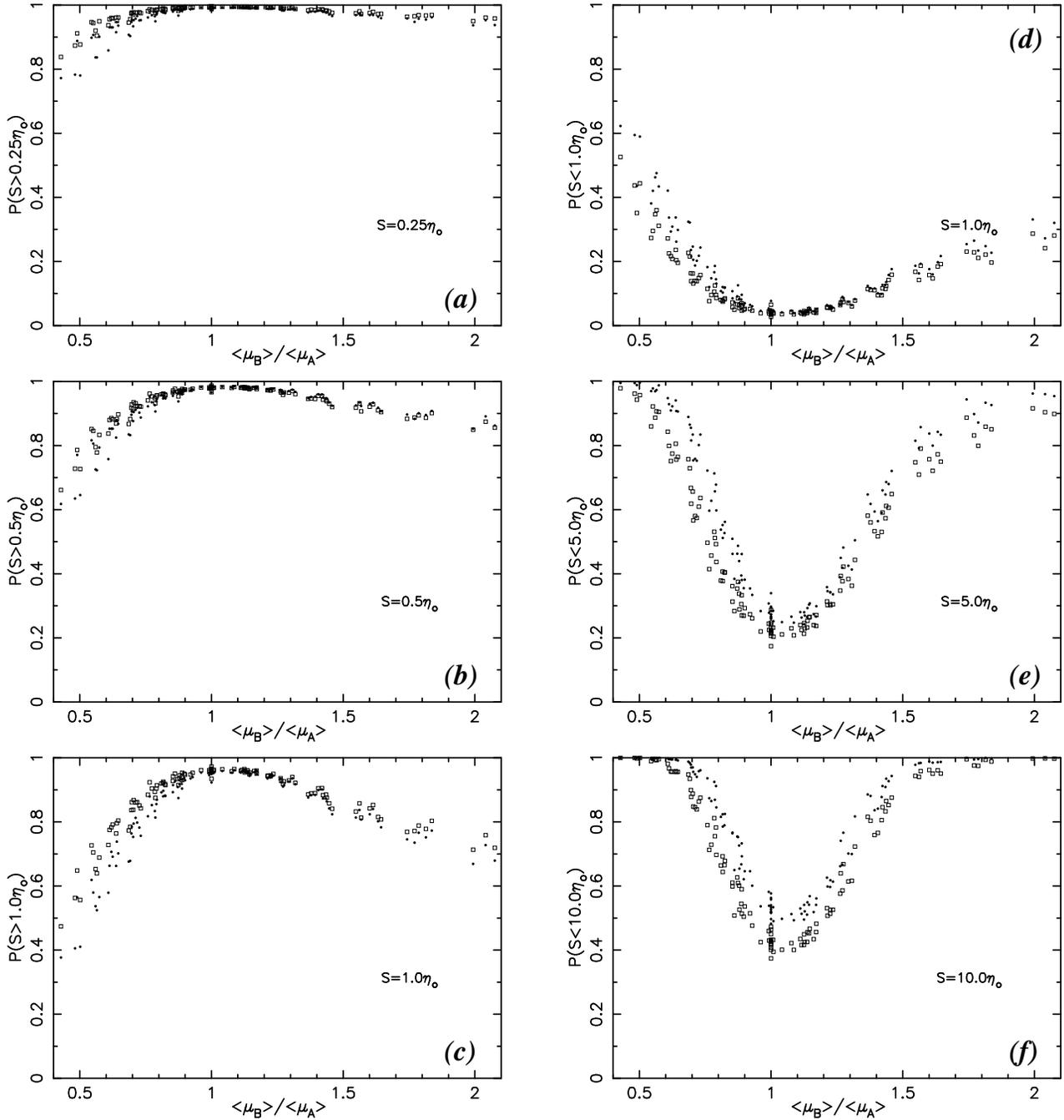}
\caption{\label{proflimits} Plots of source size limits as a function of mean flux ratio for different profiles. Panels (a)-(c): Lower limits ($S_{IR}>$0.25, 0.5 and 1.0$\eta_o$). Panels (d)-(f): Upper limits ($S_{IR}<$1.0, 5. and 10.0$\eta_o$). The dots show limits for a circular disc source of radius $S_{IR}$. The squares show results for an annular source with $\frac{S_{out}}{S_{in}}=2$ and $S_{out} = \frac{\sqrt{3}}{2}S_{IR}$. The optical source was uniform with $S_{OPT}=0.025\eta_o$. Note that panel (d) = one minus panel (c).}
\end{figure*}

The above results assumed a Gaussian mid-IR source profile. However, the results should be dependent on the source profile assumed. If the mid-IR emission is due to dust, then a torus described by an annular profile may be a more appropriate choice since there is a minimum radius inside which the dust cannot exist (e.g. AJB00). We have repeated the calculation of Eqn.~\ref{size_lim} for all combinations of macro-models $i-xi$ using circular top-hat profiles of radius $S_{IR}$, and annular profiles with outer to inner radii ratios $\frac{S_{out}}{S_{in}}=2$ and $S_{out} = \frac{\sqrt{3}}{2}S_{IR}$ (the factor of $\frac{\sqrt{3}}{2}$ preserves flux for a uniformly bright source). The left-hand panels of Fig.~\ref{proflimits} show the resulting probabilities that $S_{IR}>0.25\eta_o$, $S_{IR}>0.5\eta_o$ and $S_{IR}>1.0\eta_o$ as a function of the macro-model flux ratio $R_{BA}$. Similarly, the right-hand 3 panels of Fig.~\ref{proflimits} show the confidences that $S_{IR}<1.0\eta_o$, $S_{IR}<5.0\eta_o$ and $S_{IR}<10.0\eta_o$. The circular source limits are denoted by the dots, and the annular source limits by the squares. The consistency of the results for annular and circular profiles (particularly near $\langle\mu_B\rangle/\langle\mu_A\rangle=1$) suggests that the area of caustic structure covered by the projected source profile is the important quantity. This idea should extend to annuli of different inner to outer radii, and to the projected area of an inclined source.

\subsection{Other image combinations}
\label{sec_pairs}

\begin{figure*}
\vspace*{180mm}
\includegraphics{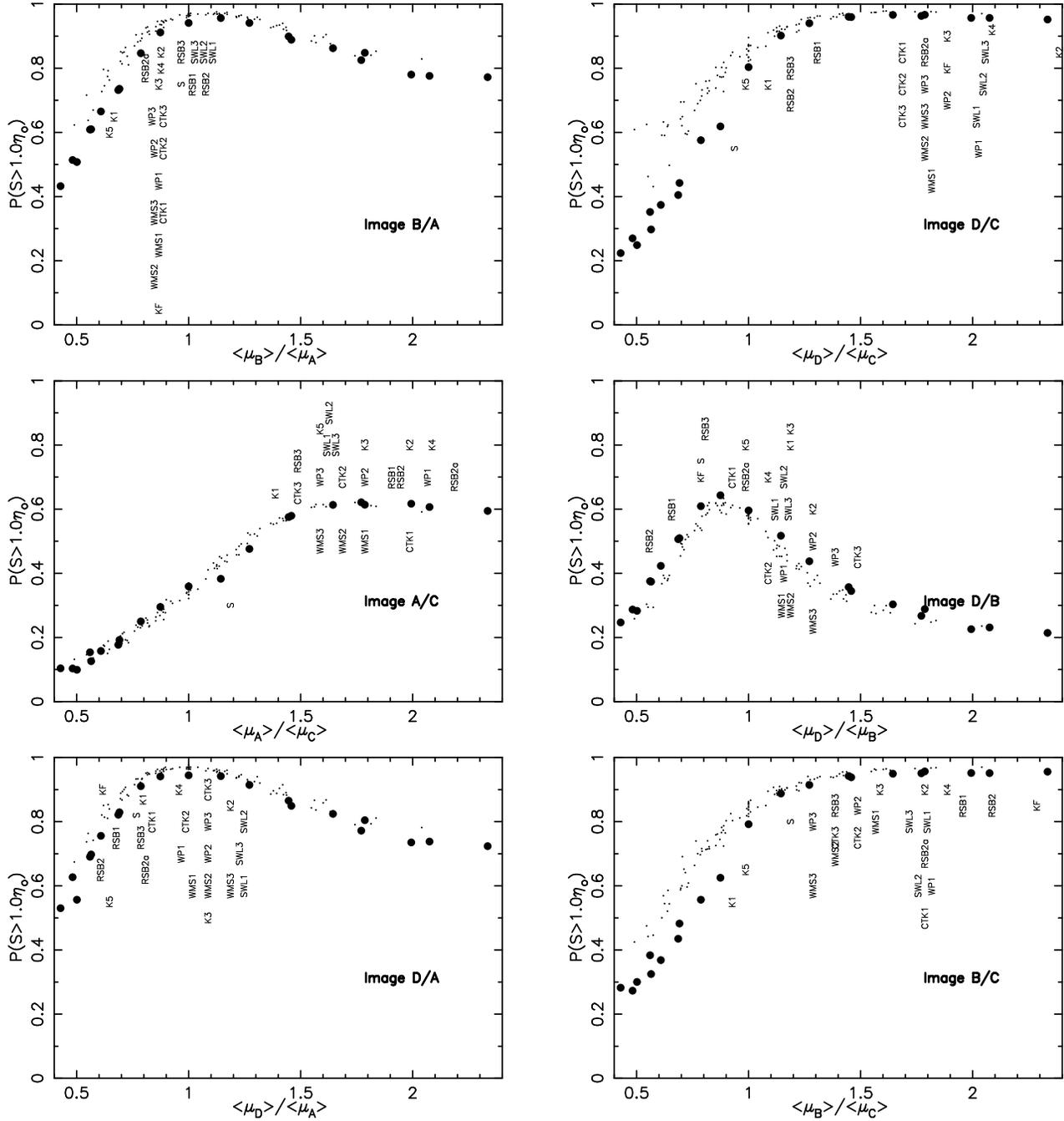}
\caption{\label{pairs} Plots of source size limits $P(S_{IR}>1\eta_o)$ as a function of mean flux ratio for each image pair. The optical source was uniform with $S_{OPT}=0.025\eta_o$. The larger markers indicate models that include the large $\kappa$ large $\gamma$ for one of the images (model $xi$). The labels correspond to published macro-models (see Tab.~\ref{macromodels}) and are plotted at the appropriate $R_{AB}$.}
\end{figure*}

Figs.~\ref{limits_single}, \ref{limits} and \ref{proflimits} show a correlation between the confidence on mid-IR source size limits and $R_{BA}$. This correlation was found for macro-models having $\kappa$ and $\gamma$ near the values found for images A and B by SWL98, but is also followed by the  macro model parameters for image C (SWL98), which has values of $\kappa$ and $\gamma$ that differ from image A by a factor of $\sim2$. This combined with the range of image magnifications over which the correlation holds leads us to assume that our ensemble of models can be used to describe the statistics for the other images as well. In particular, we use the ensemble to compute probabilities from Eqn.~\ref{size_lim} for the other 5 possible image combinations. Fig.~\ref{pairs} shows the resulting confidence that $S_{IR}>1.0\eta_o$ as a function of macro-model flux ratio $R_{ij}$. The image flux ratios used were $R_{BA}^{obs}$, $R_{AC}^{obs}$, $R_{DA}^{obs}$, $R_{BC}^{obs}$, $R_{DB}^{obs}$ and $R_{DC}^{obs}$, chosen so that $R_{ij}^{OPT,obs}<R_{ij}^{IR,obs}$ (note that only three of these ratios are independent). We used the optical fluxes from the 26th of September 1999. The predicted $R_{ij}$ of published macro-models (Tab.~\ref{macromodels}) are marked on Fig.~\ref{pairs}.

Strong limits on $S_{IR}$ are obtained from the flux ratios between images D and C, and B and C, as well as images B and A. In the former two cases, this is because all macro models predict $R>1$. In the latter case, images B and A yield the greatest agreement for $R_{ij}$ between different macro-models. These image ratios yield $S_{IR}>1\eta_o$ with $>\sim80\%$ confidence for all macro-models. $R_{BA}$ and $R_{DC}$ are independent ratios, and so their limits can be combined to produce a stronger limit of $S_{IR}>1\eta_o$ with $>\sim95\%$ confidence for all macro-models.

\section{Conclusion}

Monitoring of the gravitationally microlensed Quasar Q2237+0305 has found significant microlensed optical variability. Observations in July and September of 1999 showed (de-reddened) V-band flux ratios between the different sets of images that differed by large factors from the corresponding flux ratios in the mid-IR. In addition, the mid-IR flux ratios are similar to those measured in the radio, and to predictions of some lensing models, suggesting that mid-IR emission region is not subject to large microlensing variation. The mid-IR emission region is therefore larger than the microlens Einstein Radius ($\eta_o$), and hence at least 2 orders of magnitude larger than the optical emission region, which is thought to be $<0.01\eta_o$ (Wambsganss, Paczynski \& Schneider 1990; Wyithe, Webster \& Turner 2000). The colour difference between different images is due to the magnification/de-magnification of the optical emission. We have used microlensing models to calculate, as a function of mid-IR source size, the probability of obtaining the observed mid-IR flux ratios given the observed optical flux ratios. 

An alternative approach might be to note that the V-band flux in some images varied between the two mid-IR observations, while the mid-IR flux remained steady. In this case, the optical ratios need not be de-reddened, and as a result carry smaller errors. However, the variation in V-band is not significantly larger than the observational uncertainty in the mid-IR. We have therefore restricted our attention to flux ratio variation between different images. 

The flux ratio statistics for large sources are sensitive to the macro-model assumed. We have therefore computed source size limits for an ensemble of models. The probability has the interesting property of being primarily sensitive to the macro-model flux ratio. As a result we have parameterised the macro-models available from the literature by their predicted flux-ratio. The strongest limits are found from the flux ratios between images B and A. This is due in part to the differences between the relative V-band and mid-IR fluxes, as well as the more consistent macro-model predictions for these images. We find that the mid-IR source size $S_{IR}>1\eta_o$ with $>90\%$ confidence, and $>0.5\eta_o$ with $>95\%$ confidence. The IR-emission scale is larger than the optical emission with a confidence $>99\%$.

The limit on the infrared source size derived here for the Einstein Cross may be converted to a limit on the brightness temperature.  Assuming microlensing by stars, $S_{IR} \ga \eta_o \sim 10^{17}$cm, and the brightness temperature at 10 $\mu$m (rest frame 3.7 $\mu$m) is about $T_b \la 7900 K$ for a luminosity distance of
$10^{28} $cm, magnification of 15, and flux of 20 mJy.  This
upper limit on the source brightness rules out non-thermal emission mechanisms,
such as synchrotron, which typically have brightness temperatures of $10^{8-10}$K. As argued in AJB00, the spectrum indicates thermal emission at $\sim 2000$K,
which is comparable to the sublimation temperature of dust.   The flux from
the QSO is sufficient to heat the dust at the sublimation radius of $\sim$~1~pc.

One can turn this argument around: assuming that the IR emission is due to dust
at the sublimation radius, then one can estimate the variance of the
flux of each image for the given source size and macro lens model. We can 
then compute the $\chi^2$ of each macrolensing model using this variance.
Analytic estimates of the variance for large source exist for models with
$\gamma=0$ \cite{R93} and some numerical estimates exist for non-zero
$\gamma$ \cite{R97}.  The estimates with zero $\gamma$ indicate that the
variance should be of order 10\% or less, unfortunately about the same size 
as the observational error bars for the infrared observations.  Since models 
for a large source size require much larger ray-tracing simulations, with a lower spatial resolution than we
have  carried out, we leave this computation for future work.  Future 
observations with higher signal-to-noise, or a measurement of the variance
with long-term IR monitoring will allow a better estimate of which macro-lens
model is correct.

\section*{Acknowledgements}

The authors would like to thank Rachel Webster, Ed Turner and Joachim Wambsganss for very helpful discussions and comments on the manuscript. This work was supported in part by NSF grant AST120-6217 to Ed Turner. JSBW acknowledges the support of an Australian Postgraduate Award and a Melbourne University Overseas Research Experience Award. The simulations presented in this paper were performed at the Centre for Astrophysics and Supercomputing, Swinburne University of Technology.

\label{lastpage}

\end{document}